# A variation of the clock paradox and a distinguishing feature of an inertial frame

Chandru Iyer, Techink Industries, C-42, phase-II, Noida, 201305, India, chandru_i@yahoo.com
G. M. Prabhu, Dept. of Computer Science, ISU, Ames, IA 50011, USA, prabhu@cs.iastate.edu

**Abstract**

The clock paradox is analyzed for the case when the onward and return trips cover the same 'distance' (as observed by the traveling twin) but at unequal velocities. In this case the stationary twin observes the distances covered by her sister during the onward and return trips to be different. The analysis is presented using formulations of special relativity and the only requirement for consistency is that all observations are made from any one chosen inertial frame. The analysis suggests that a defining feature of an inertial frame should be based on the continued maintenance of the distinctive synchronicity of the clocks co-moving with it.

## 1. Introduction

The clock or twin paradox is a widely discussed topic in relativity, see for example [1, 2, 3]. In one formulation, the problem concerns two twins, Eartha and Stella, who are present together in an inertial frame. After initial synchronization of their clocks, Eartha stays stationary on earth, while Stella undertakes a round trip. In the onward journey she travels for 5 years at a velocity of +0.8$c$; in the return journey she travels for 5 years at a velocity of −0.8$c$ (these observations are made from Eartha's frame). When Stella returns the twins compare their clocks and find that Eartha's reads 10 years, whereas Stella's reads 6 years. Since all physical processes also follow the same logistics of the clocks they carry, Eartha finds herself older by 10 years and Stella finds herself older by six years. Thus Stella is younger than Eartha by 4 years. Except for the brief periods when Stella was accelerating, the twins belong to inertial frames and therefore, by the assertion of special relativity, the situation should appear perfectly symmetrical and can be analyzed from either frame, considering a state of rest for that frame. Thus Stella can consider herself at rest and would find Eartha to be younger than her after the round trip. This generates the paradox because permanent changes in aging should be observer-independent and each clock cannot be behind the other.

Apparently the twin paradox was presented as a thought problem in relativity by P. Langevin in 1911 [4], although it may have prior history [5]. Einstein addressed the twin paradox in special relativity in 1918 in the form of a dialogue between a critic and a relativist [6]. Einstein admitted that the special relativistic time dilation was symmetric for the twins and he had to invoke, asymmetrically, the general relativistic time dilation during the brief periods of acceleration to justify the asymmetrical aging. The role of acceleration has been examined in detail in [7–10] and a full general relativistic treatment has been given in [11].

The formulations of special relativity decree that moving clocks and physical processes 'appear' to run slow by a factor $\sqrt{1 - v^2/c^2}$, where $v$ is the relative velocity between the clock and a chosen inertial frame. For a co-moving inertial frame (with the clock), no such slow running of the clock (or physical process) is observable. In most of the standard resolutions of the paradox, it is assumed that the correct calculation is to be done from Eartha's inertial frame because Stella accelerated and broke the symmetry of inertial motion. The space-time diagram is drawn with the time in Eartha's frame as the vertical axis and Stella's world line indicated as being shorter in the Lorentzian sense. A space-time diagram with Stella's proper time as the vertical axis leads to opposite conclusions and is avoided because such a calculation is deemed beyond the scope of special relativity. If Stella's onward velocity is taken to be $c\beta$, where $c$ is the speed of light, then Stella's onward travel time $\Delta t_1$ as measured by her is related to Eartha's measurement $\Delta t_2$ of the same interval by the relation

$$\Delta t_1 = \Delta t_2 \sqrt{1-\beta^2} \,. \tag{1}$$

In the above relation, Stella measures the onward journey travel time by her particular clock. The particular clock carried by Eartha cannot be used by her to measure the time at which Stella made the actual turnaround, because there is a considerable spatial separation between them when this event occurs. In order for Eartha to measure the time interval of Stella's onward journey, she needs to have a clock positioned at the spatial location of the actual turn around point of Stella. Needless to say, this clock should be synchronized with the particular clock carried by Eartha, in order for the time interval of Stella's onward journey to be correctly measured by Eartha.

The formulations of special relativity decree (by symmetry) that the relation

$$\Delta t_2 = \Delta t_1 \sqrt{1-\beta^2} \tag{2}$$

is also valid where $\Delta t_2$ is the time elapsed in the particular clock with Eartha during the onward journey of duration $\Delta t_1$ as observed by Stella.

The equations (1) and (2) above, even though symmetrical, are apparently contradictory because multiplying the left hand and right hand sides of the two equations and equating the results yields $\beta = 0$, which negates the relative motion between the two inertial frames. However, this contradiction can be resolved by realizing that in equation (1) Stella's inertial frame measures the time interval between the two events (start and finish of Stella's onward journey) by a particular clock, whereas Eartha's inertial frame measures the same time interval by spatially separated clocks. In equation (2) the time elapsed by the particular clock carried by Eartha is determined by Stella using spatially separated clocks in Stella's inertial frame. Similar relations can be formulated for Stella's return journey. According to these relations, Eartha finds Stella younger when she returns and Stella finds Eartha younger when she returns.

The twin paradox may be analyzed under special relativity from *three* reference frames, namely, Eartha's (E), Stella's onward journey frame ($S_1$), and Stella's return journey frame ($S_2$). We designate the time interval between the turning around of Stella and the initial departure of Stella as $T_1$ and the time interval between the return of Stella and the turning around of Stella as $T_2$. The observed values of $T_1$ and $T_2$ are different in the three inertial frames E, $S_1$ and $S_2$. But they all agree on the total times clocked by the particular clocks carried by Eartha ($C_E$) and Stella ($C_S$) as detailed below:

Frame E:   $T_1$ = 5 years;         Clocked by $C_E$ = 5 years;    Clocked by $C_S$ = 3 years
           $T_2$ = 5 years;         Clocked by $C_E$ = 5 years;    Clocked by $C_S$ = 3 years

Frame $S_1$:  $T_1$ = 3 years;          Clocked by $C_E$ = 1.8 years; Clocked by $C_S$ = 3 years
              $T_2$ = 13.667 years;  Clocked by $C_E$ = 8.2 years; Clocked by $C_S$ = 3 years

Frame $S_2$:  $T_1$ = 13.667 years;  Clocked by $C_E$ = 8.2 years; Clocked by $C_S$ = 3 years
              $T_2$ = 3 years;           Clocked by $C_E$ = 1.8 years; Clocked by $C_S$ = 3 years

(Explanation of the above numerical values:

Frame E (Eartha's reference frame) observes Stella's clock to be running slow both during her onward and return journeys by a factor 0.6 corresponding to the relative velocity of 0.8$c$ between Stella and Eartha.

Frame $S_1$ observes Eartha's clock to be running slow. During the onward journey, this is the inertial frame in which Stella is present. Since Stella clocks 3 years during the onward journey, Eartha clocks 1.8 years as observed by frame $S_1$. During the return journey, Stella switches to frame $S_2$ and the frame $S_1$ becomes an abstract frame. As observed by frame $S_1$, during the return journey, Stella's particular clock (which is in frame $S_2$) clocks 3 years but is running slow by a factor 9/41 corresponding to the relative velocity between $S_1$ and $S_2$ of 40$c$/41. Therefore, the actual time duration of the return journey, as observed by frame $S_1$, is 13.667 years. Frame $S_1$ observes Eartha's clock to run slow by a factor 0.6 corresponding to the relative velocity between frame $S_1$ and E of 0.8$c$. Therefore, during the return journey, Eartha clocks 13.667 * 0.6 = 8.2 years, as observed by frame $S_1$.

Observations of frame $S_2$ are very similar to that of frame $S_1$; only the two legs of Stella's journey are interchanged.)

Thus the total time clocked by $C_E$ and $C_S$ are observed to be 10 years and 6 years, respectively, by all the three inertial frames E, $S_1$ and $S_2$ and in fact by any arbitrary inertial frame.

The twin paradox may also be analyzed by invoking the general relativistic time dilation during the brief periods of Stella's acceleration [12]. Eartha observes that Stella is younger because of Stella's kinematical velocity, and Stella observes that Eartha is older because clocks run faster under a gravitational potential. Under the general relativistic formulations explained in [12], Eartha ages by 6.4 years (= 8.2 – 1.8) when Stella turns around. According to Eartha, Stella's clock ran slow during both the onward and return journeys and this is the reason that Stella is younger than her. According to Stella, Eartha's clocks ran slower during the onward and return journeys and by this consideration Eartha is younger by 2.4 years. However, Eartha aged an additional 6.4 years during the turn around and therefore Stella is younger by 4 years. Thus the twins agree on the end result but have a different explanation for it. An interesting version of the twin paradox without acceleration has been described in [13]. In this case, the twins meet again without either of them being accelerated. This was possible because of the helical path one of the twins took in a cylindrical space-time continuum. The author in [13] concludes that the world lines of the two paths embedded in the space-time continuum are not symmetric even though both the observers did not experience any acceleration; in order for two observers to agree on elapsed time between meetings, it is not sufficient that their motion is symmetrical in terms of acceleration felt; their world lines embedded in the space-time continuum must also be symmetric [13].

In this paper we consider a variation of the twin paradox problem by choosing different values for the onward and return velocities. Stella travels for 3 years (according to Stella's clock) at a velocity of 0.8$c$ with respect to Eartha. Then Stella turns around and travels at a velocity of –0.6$c$ for 4 years (according to Stella's clock). We show that although Stella covers 2.4 ly (light years) on both the onward and return journeys, she does not return to the starting point; that is, Stella does not get to meet Eartha at the end of the return journey.

The situation is resolved by invoking the principle that all observations must be made from the same inertial frame in order to arrive at correct conclusions. Stella traveled 2.4 light years during her onward journey as observed by the inertial frame co-moving with her during the onward journey. Similarly, she traveled –2.4 ly during her return journey as observed by the inertial frame co-moving with her during her return journey. Since these two observations are made from two different inertial frames, they cannot be conflated to reach a correct conclusion.

From a mathematical standpoint, the variation we have considered may not appear significant. However, from the point of view of the traveling twin (Stella), the physical situation presents an additional paradox as baffling as the conventional clock paradox. Stella trusts her clocks and measuring devices. According to Stella's observations, Eartha traveled at –0.8$c$ for 3 years and +0.6$c$ for 4 years, as measured by Stella's own measuring devices. Therefore Stella expected to meet Eartha at the end of the journey. When Eartha and Stella do not meet, the situation makes Stella wonder which of her measurements are in error. For Eartha it is not a paradox, because she observes that Stella traveled at +0.8$c$ for 5 years and –0.6$c$ for 5 years, and therefore Eartha does not expect to meet Stella.

Even in the Galilean/Newtonian dispensations the above principle that 'all observations must be made from the same frame' is very much valid. As in the case of two objects, A and B of mass $m_1$ and $m_2$ respectively, moving with respect to each other at a relative velocity of ± u, an observer present in the co-moving inertial frame of object A, will observe the total kinetic energy of the system to be ½ $m_2 u^2$ and the total momentum of the system to be $m_2 u$. Whereas an observer present in the co-moving inertial frame of object B, will observe these quantities to be ½ $m_1 u^2$ and $m_1 u$ respectively. Suppose an observer switches from one inertial frame to another inertial frame. After switching frames, this observer cannot draw correct conclusions based on the data from the previous inertial frame in which he was present, in combination with the data obtained as observed from his current inertial frame.

We show that observations made from any of the three inertial frames (i) Eartha's frame E, or (ii) Stella's onward journey frame $S_1$, or (iii) Stella's return journey frame $S_2$ yield consistent results. There is no contradiction between the observations of the three frames under the principles of special relativity. We further show that observations from any arbitrary inertial frame K moving at a velocity $c\beta$ with respect to Eartha also yield consistent results.

Many studies [11, 12] consider the problem from the view point of Stella who considers herself stationary and observes Eartha having traveled and returned. However, when Eartha and Stella meet for the second time, it is Stella's clock that shows a smaller time interval. We postulate that the distinguishing feature between Eartha's inertial frame and Stella's reference frame is not "Who traveled" but which of the two was able to maintain a consistency in the synchronicity of the clocks co-moving with her. It is this feature, we suggest, that distinguishes an inertial frame from a non-inertial frame, rather than uniform velocity, because the observation of a non-uniform velocity between two frames does not allow one to infer the non-uniformity of motion of either frame.

## 2. Observations from Eartha's co-moving frame (E)

Eartha's frame (E) is the "stationary" frame in which both twins are originally present. It is observed from this frame that Stella's onward journey lasts 5 years at $v$ = 0.8$c$; but Stella's clock times it as $5\sqrt{1-(0.8)^2} = 3$ years (because of time dilation as the moving twin can be considered as a moving clock). From Eartha's frame, Stella's return journey also takes 5 years at $v$ = 0.6$c$; but Stella's clock times it as $5\sqrt{1-(0.6)^2} = 4$ years. In the onward journey Stella travels $5 \times 0.8 = 4$ ly and in the return journey Stella travels $5 \times 0.6 = 3$ ly.

Thus at the end of the return journey Stella does not reach Eartha and the distance between them is 1 light year. In order for Stella to reach Eartha, the return journey duration should be extended to 4/0.6 = 20/3 years (as measured by Eartha's clock), and 0.8(4/0.6) = 16/3 years (as measured by Stella's clock).

### 3. Observations from Stella's co-moving onward journey frame (S₁)

With respect to Stella's onward journey frame $S_1$, Stella is stationary during the 'onward journey' and the duration of this phase is 3 years. Eartha is moving at a velocity of –0.8$c$ with respect to $S_1$ during this phase.

During the second phase (return journey) Stella travels at $\frac{(-0.8 - 0.6)c}{1 + (0.8 * 0.6)} = \frac{-35c}{37}$ with respect to $S_1$ and Eartha is still moving at –0.8$c$. Even though Stella's clock measures the duration as 4 years, the actual time according to the inertial frame $S_1$ is $\frac{4}{\sqrt{1 - \frac{35^2}{37^2}}} = \frac{37}{3}$ years. The approach velocity between Stella and Eartha during the second phase is $\frac{35c}{37} - 0.8c = \frac{54c}{370}$. The distance of separation between Eartha and Stella at the end of the first phase is –2.4 light years (= 3 * (–0.8$c$)).

The distance covered during the second phase is $\frac{37}{3} \times \frac{54c}{370} = \frac{54}{30}c$ = 1.8 light years. Therefore, there is still a separation of 0.6 light years left between Stella and Eartha. In Eartha's co-moving frame (E), this separation will be observed as $\frac{0.6}{\sqrt{1 - (0.8)^2}} = 1$ ly, which is consistent with observations described in Section 2.

### 4. Observations from Stella's co-moving return journey frame (S₂)

From Stella's return journey frame $S_2$, the onward journey frame $S_1$ is moving at a velocity of $+\frac{35c}{37}$ and Eartha is moving at a velocity of +0.6$c$. So during the onward journey, even though Stella clocked 3 years by her clock, the time duration was actually $\frac{3}{\sqrt{1 - \frac{35^2}{37^2}}} = \frac{37}{4}$ years. Stella and Eartha got separated by $\frac{37}{4}[\frac{35}{37}c - 0.6c] = 3.2$ ly.

During the return journey, according to frame $S_2$, Stella was 'stationary' for 4 years and Eartha moved at 0.6$c$, thereby covering a distance of 4 × 0.6 = 2.4 ly. Thus the distance remaining between Eartha and Stella is [3.2 – 2.4] ly = 0.8 ly. This distance of 0.8 ly will be observed by Eartha's frame E as $\frac{0.8}{\sqrt{1 - (0.6)^2}} = 1$ ly, which again is consistent with the observations in Sections 2 and 3.

## 5. Observations from an inertial frame K moving at *cβ* with respect to Eartha's co-moving inertial frame E

From frame K, the velocity of the onward journey is $(\dfrac{0.8-\beta}{1-0.8\beta})c$, which can be expressed as $c\beta_1$ and the velocity of the return journey is $(\dfrac{0.6+\beta}{1+0.6\beta})c$, which can be expressed as $c\beta_2$.

The actual time duration of the onward journey is $\dfrac{3}{\sqrt{1-\beta_1^2}}$ years, and the separation between Eartha and Stella during this journey is $\dfrac{3c}{\sqrt{1-\beta_1^2}}(\beta_1+\beta)$ light years. During the return journey, the distance covered is $\dfrac{4c}{\sqrt{1-\beta_2^2}}(\beta_2-\beta)$ light years. Thus the net distance in light years at the end of the return journey is given by

$$\dfrac{3c}{\sqrt{1-\beta_1^2}}(\beta_1+\beta) - \dfrac{4c}{\sqrt{1-\beta_2^2}}(\beta_2-\beta). \tag{3}$$

By straightforward algebraic simplification, equation (3) reduces to $c\sqrt{1-\beta^2}$ after substituting for $\beta_1$, $\beta_2$ as $(\dfrac{0.8-\beta}{1-0.8\beta})$ and $(\dfrac{0.6+\beta}{1+0.6\beta})$ respectively (proof given below).

**Proof.** To prove that $\dfrac{3(\beta_1+\beta)}{\sqrt{1-\beta_1^2}} - \dfrac{4(\beta_2-\beta)}{\sqrt{1-\beta_2^2}} = \sqrt{1-\beta^2}$.

$$\text{LHS} = \dfrac{3(\dfrac{0.8-\beta}{1-0.8\beta}+\beta)}{\sqrt{1-\dfrac{(0.8-\beta)^2}{(1-0.8\beta)^2}}} - \dfrac{4(\dfrac{0.6+\beta}{1+0.6\beta}-\beta)}{\sqrt{1-\dfrac{(0.6+\beta)^2}{(1+0.6\beta)^2}}}$$

$$= \dfrac{3(0.8-0.8\beta^2)}{0.6\sqrt{1-\beta^2}} - \dfrac{4}{0.8}\dfrac{(0.6-0.6\beta^2)}{\sqrt{1-\beta^2}}$$

$$= \dfrac{5}{\sqrt{1-\beta^2}}[0.8(1-\beta^2) - 0.6(1-\beta^2)]$$

$$= (5*0.2)\sqrt{1-\beta^2} = \sqrt{1-\beta^2} = \text{RHS} \blacksquare$$

Thus at the end of the return journey, the separation between Eartha and Stella is $\sqrt{1-\beta^2}$ light years.

## 6. Summary

We have shown in this paper that from an arbitrarily chosen inertial frame (K), the distance of separation between Eartha and Stella is $\sqrt{1-\beta^2}$ light years at the end of Stella's onward and return journeys. In Eartha's co-moving frame, for the instance described in this paper, the separation distance is measured as 1 light year. When $\beta = 0$, frames K and E become identical and observations from frames K and E are the same. When $\beta = 0.8$, observations from frames K and $S_1$ are identical. When $\beta = -0.6$, observations from frames K and $S_2$ are identical. We have analyzed this problem under special relativity, focusing on the three inertial frames relevant to the physical situation and extending it to any arbitrary inertial frame. Stella's reference frame (S = $S_1$ + $S_2$) is considered as a legitimate single reference frame under general relativity. Therefore, when the problem is analyzed from Stella's point of view (as observed by reference frame S), the use of general relativistic considerations become imperative.

All inertial frames observe that moving clocks run slow. However, when they meet for the second time, Stella observes that Eartha's clock has run faster. A point in question is: What made Stella's clock run slow compared to Eartha's? Or what distinguishes Eartha and Stella, when both observed each other to be in relative motion throughout the entire episode? An important distinction is that while Eartha was able to generate a set of synchronized clocks (that are not universal but with a synchronicity unique to Eartha) Stella would have found that this was not possible for her as the synchronicity of the clocks co-moving with Stella would be different before and after her "turnaround." This implies that a reference frame can be classified as "inertial" or "non-inertial" by its ability or inability to create a consistent synchronicity of clocks co-moving with it. This means an acceptable process that creates a set of spatially separated yet synchronous clocks, when repeated in an inertial frame after a time interval finds that the synchronicity of the co-moving clocks is unaffected and there is no need to re-synchronize the clocks.

A frame of reference that can create a consistent set of synchronized clocks (unique to that frame) which are co-moving with it can be classified as "inertial." Our definition differs from the standard definition of the inertial frame as one which moves with "uniform velocity." This change in definition is proposed because if two reference frames observe each other to be moving at a non-uniform velocity, it is difficult to say which one of them is actually moving at a non-uniform velocity. But with our new definition, the frame that is able to maintain the consistency of synchronicity among spatially separated clocks co-moving with it, will qualify to be an "inertial frame." In other words, for a reference frame, the time interval during which the synchronicity of co-moving clocks is affected, defines the period when the frame is non-inertial.

### Acknowledgements

The authors would like to thank the reviewers for their comments. This paper has benefited from their careful work.